\documentclass[3p,times,twocolumn]{elsarticle}

\usepackage{ecrc}

\volume{00}

\firstpage{1}

\journalname{Nuclear Physics B Proceedings Supplement}

\runauth{X.-Q. Li, J. Lu and A. Pich}

\jid{nuphbp}

\jnltitlelogo{Nuclear Physics B Proceedings Supplement}

\usepackage{amssymb}
\usepackage[figuresright]{rotating}
\usepackage{caption}
\usepackage{amssymb,amsmath}

\usepackage{graphicx} 
\usepackage{subfigure}

\newcommand{\cR}{\mathcal{R}}

\newcommand{\Bqll}{B_{s,d}^0 \to \ell^+\ell^-}

\newcommand{\Bsmm}{B_s^0 \to \mu^+\mu^-}
\newcommand{\Bdmm}{B_d^0 \to \mu^+\mu^-}
\newcommand{\nn}{\nonumber}

\def\be{\begin{equation}}
\def\ee{\end{equation}}
\def\beqn{\begin{eqnarray}}
\def\eeqn{\end{eqnarray}}
\def\no{\nonumber}
\def\ba{\begin{array}{c}}
\def\bat{\begin{array}{cc}}
\def\ea{\end{array}}
\def\bi{\begin{itemize}}
\def\ei{\end{itemize}}

\def\cC{{\cal C}}

\begin{document}

\begin{frontmatter}



\dochead{}

\title{$\Bqll$ Decays in Two-Higgs Doublet Models}
%\title{$\Bqll$ Decays in the Aligned Two-Higgs Doublet Model}


\author{Xin-Qiang Li}
\address{Institute of Particle Physics and Key Laboratory of Quark and Lepton Physics~(MOE),
\\ Central China Normal University, Wuhan, Hubei 430079, P.~R.~China}
\address{State Key Laboratory of Theoretical Physics, Institute of Theoretical Physics,\\
 Chinese Academy of Sciences, Beijing 100190, P.~R.~China}

\author{Jie Lu \fnref{lj}}\fntext[lj]{Speaker}
\author{Antonio Pich}
\address{IFIC, Universitat de Val\`encia -- CSIC, Apt. Correus 22085, E-46071 Val\`encia, Spain}

\begin{abstract}
We study the rare leptonic decays $\Bqll$ within the general framework of the aligned two-Higgs doublet model~\cite{Li:2014fea}. A complete one-loop calculation of the relevant short-distance Wilson coefficients is presented, with a detailed technical summary of the results. The phenomenological constraints imposed by present data on the model parameters are also investigated.
% We also compare our results with previous calculations performed in particular limits or approximations, and investigate the impact of the model parameters on the branching ratios.
\end{abstract}


\end{frontmatter}

% \linenumbers

\section{Introduction}
\label{sec:intro}

The discovery~\cite{Aad:2012tfa,Chatrchyan:2012ufa} of a Higgs-like boson at the LHC has placed the last missing piece of the Standard Model~(SM), which is one of the greatest achievements of modern particle physics. However, it is widely believed that the SM cannot be the fundamental theory up to the Plank scale, and many theories beyond the SM~(BSM) claim that new physics (NP) should appear around the TeV scale.

One of the simplest extensions of the SM is the addition of an extra Higgs doublet~\cite{Lee:1973iz}. Two scalar doublets are present in several BSM theories, for instance in supersymmetry.
% Lagrangian~\cite{Lee:1973iz}. The requirement of two-Higgs doublets also exists in many more fundamental BSM theories, such as super-symmetry, left-right symmetry, little Higgs, etc..
%
Two-Higgs doublet models (2HDMs) with generic Yukawa couplings give rise to dangerous
tree-level flavour-changing neutral currents~(FCNCs) \cite{2HDM:review}.
%usually poses the problem of tree-level flavour-changing neutral current~(FCNC), which is forbidden within the SM.
This can be avoided imposing discrete $Z_2$ symmetries \cite{Glashow:1976nt} or, more generally, assuming
% by imposing discrete $\mathcal{Z}_2$ symmetry in the Higgs sector. A more general alternative is, on the other hand, to assume
the alignment in flavour space of the two Yukawa matrices for each type of right-handed fermions~\cite{Pich:2009sp}.
%, which is named as the aligned two-Higgs doublet model~(A2HDM).

The leptonic decays $\Bqll$ play a very special role in testing the SM and probing BSM physics. % These processes
They are very sensitive to the mechanism of quark flavour mixing, and their branching ratios are extremely small due to the loop suppression and the helicity suppression factor $m_{\ell}/m_b$. Since the final state involves only leptons, the SM theoretical predictions
%for these decay modes
are very clean~\cite{Bobeth:2013uxa}:
%, with the latest predictions given as~\cite{Bobeth:2013uxa}:
\begin{align} \label{eq:BqmmSM}
\overline{\mathcal{B}}(\Bsmm) &= (3.65 \pm 0.23) \times 10^{-9}\,, \\
\overline{\mathcal{B}}(\Bdmm) &= (1.06 \pm 0.09) \times 10^{-10}\,,
\end{align}
which include
%where the
next-to-leading order~(NLO) electroweak corrections~\cite{Bobeth:2013tba} and
%the
next-to-next-to-leading order~(NNLO) QCD corrections~\cite{Hermann:2013kca}.
% have been taken into account.

The weighted world averages of the CMS~\cite{Chatrchyan:2013bka} and LHCb~\cite{Aaij:2013aka} measurements~\cite{CMSandLHCbCollaborations:2013pla}
\begin{align} \label{eq:Bsmmexp}
\overline{\mathcal{B}}(\Bsmm)_{\rm exp.} &=  (2.9 \pm 0.7) \times 10^{-9}\,,\\
\overline{\mathcal{B}}(\Bdmm)_{\rm exp.} &= \left(3.6^{\, +1.6}_{\, -1.4}\right) \times 10^{-10}\, ,
\end{align}
are very close to the SM predictions and put stringent constraints on BSM physics.

\section{The aligned two-Higgs doublet model}
\label{sec:A2HDM}

%\subsection{The scalar sector}

%The 2HDM is often defined in the ``Higgs basis", especially for the Yukawa coupling sector. In this basis, only one doublet gets a nonzero vacuum expectation value, and the two doublets can be parametrized as
It is convenient to define the 2HDM in the ``Higgs basis" where only one scalar doublet gets a nonzero vacuum expectation value $v=(\sqrt{2} G_F)^{-1/2} \simeq 246~\mathrm{GeV}$:
\begin{align}
 \Phi_1 &= \left[ \begin{array}{c} G^+ \\ \frac{1}{\sqrt{2}}\, (v+ S_1 +i\, G^0) \end{array} \right] \; , \\
 \Phi_2 &= \left[ \begin{array}{c}  H^+ \\ \frac{1}{\sqrt{2}}\, ( S_2 +i\,  S_3 )   \end{array}\right] \; .
\end{align}
The first doublet contains the Goldstone fields $G^\pm$ and $G^0$.
%, and $v=(\sqrt{2} G_F)^{-1/2} \simeq 246~\mathrm{GeV}$.
The five physical degrees of freedom are given by the two charged fields $H^\pm(x)$ and three neutral scalars $\varphi^0_i(x) =\{h(x), H(x), A(x)\}$. The latter are related with the $S_i$ fields through an orthogonal transformation $\mathcal{R}$, which defines the neutral mass eigenstates:
\begin{equation} \label{eq:mass_diagonalization}
\mathcal{R}\,\mathcal{M}\,\mathcal{R}^T  =  \mathrm{diag}\left( M_h^2, M_H^2, M_A^2\right)\,, \qquad
\varphi^0_i = \mathcal{R}_{ij}  S_j\,.
\end{equation}
The mass matrix $\mathcal{M}$ of the neutral scalars is fixed by the scalar potential:
\begin{align}
%\hspace{-1.0cm}
V &= \mu_1\left(\Phi_1^\dagger\Phi_1\right)+ \mu_2 \left(\Phi_2^\dagger\Phi_2\right) %\nn\\[0.2cm] &
+ \left[\mu_3 \left(\Phi_1^\dagger\Phi_2\right) + \mu_3^* \left(\Phi_2^\dagger\Phi_1\right)\right] \no\\[0.2cm]
& + \lambda_1\, \left(\Phi_1^\dagger\Phi_1\right)^2  + \lambda_2 \left(\Phi_2^\dagger\Phi_2\right)^2  \nn\\
& + \lambda_3 \left(\Phi_1^\dagger\Phi_1\right) \left(\Phi_2^\dagger\Phi_2\right)  + \lambda_4 \left(\Phi_1^\dagger\Phi_2\right) \left(\Phi_2^\dagger\Phi_1\right) \\[0.2cm]
& + \left[\left(\lambda_5 \Phi_1^\dagger\Phi_2  +\,\lambda_6 \Phi_1^\dagger\Phi_1  +\,\lambda_7 \Phi_2^\dagger\Phi_2\right) \left(\Phi_1^\dagger\Phi_2\right) + \mathrm{h.c.}\right]\,,\nn
\end{align}
where $\mu_1$, $\mu_2$ and $\lambda_{1,2,3,4}$ are real, while $\mu_3$ and $\lambda_{5,6,7}$ can be complex.

%The mass matrix $\mathcal{M}$ of neutral scalars is diagonalized by an orthogonal matrix $\mathcal{R}$, which defines the neutral mass eigenstates:
%\begin{equation} \label{eq:mass_diagonalization}
%\mathcal{R}\,\mathcal{M}\,\mathcal{R}^T  =  \mathrm{diag}\left( M_h^2, M_H^2, M_A^2\right)\,, \qquad
%\varphi^0_i = \mathcal{R}_{ij}  S_j\,.
%\end{equation}
In the CP-conserving limit, the neutral Higgs spectrum contains a CP-odd field $A=S_3$ and two CP-even scalars $h$ and $H$ which mix through the two-dimensional rotation matrix:
\begin{equation} \label{eq:CPC_mixing}
\left(\ba h\\ H\ea\right)\; = \;
\left[\bat \cos{\tilde\alpha} & \sin{\tilde\alpha} \\ -\sin{\tilde\alpha} & \cos{\tilde\alpha}\ea\right]\;
\left(\ba S_1\\ S_2\ea\right)\,.
\end{equation}
We use the conventions $M_h \le M_H$, and $0 \leq \tilde\alpha \leq \pi$ so that $\sin{\tilde\alpha}$ is always positive.

\subsection{Yukawa sector}

The 2HDM Yukawa sector is given by
\begin{align} \label{eq:Yukawa1}
 \mathcal{L}_Y &= -\frac{\sqrt{2}}{v}\;
 \Big[\bar{Q}'_L (M'_d \Phi_1 + Y'_d \Phi_2) d'_R \nn \\
 & \qquad \quad + \bar{Q}'_L (M'_u \tilde{\Phi}_1 + Y'_u \tilde{\Phi}_2) u'_R  \nn \\
 & \qquad \quad + \bar{L}'_L (M'_\ell \Phi_1 + Y'_\ell \Phi_2) \ell'_R \Big]\, +\, \mathrm{h.c.} \,,
\end{align}
with $\tilde{\Phi}_i(x)=i\tau_2\Phi_i^{\ast}(x)$ the charge-conjugated scalar doublets with hypercharge $Y=-\frac{1}{2}$. $Q'_L$ and $L'_L$ denote the SM left-handed quark and lepton doublets, respectively, and $u'_R$, $d'_R$ and $\ell'_R$ are the corresponding right-handed singlets, in the weak interaction basis.

The Yukawa couplings $M'_f$ and $Y'_f$~($f=u,d,\ell$) are complex $3\times3$ matrices which, in general, cannot be diagonalized simultaneously, generating FCNCs at tree level. This can be avoided by assuming that $M'_f$ and $Y'_f$
%the two matrices
are proportional to each other~\cite{Pich:2009sp}. In the mass-eigenstate fermion basis with diagonal matrices $M_f$, one has then
%In the fermion mass-eigenstate basis with diagonal mass matrices $M_f$, we can assume that the two Yukawa coupling matrices are proportional to each other~\cite{Pich:2009sp},
\begin{equation} \label{eq:alignment}
 Y_{d,\ell} = \varsigma_{d,\ell}\, M_{d,\ell}\, ,
 \qquad
 Y_u = \varsigma^*_u\, M_u\, ,
\end{equation}
with arbitrary complex parameters $\varsigma_f$~($f=d,u,\ell$), which introduce new sources of CP violation. The aligned 2HDM (A2HDM) Yukawa Lagrangian reads
%With this alignment condition, the Lagrangian~(\ref{eq:Yukawa1}) can be rewritten as
\begin{align} \label{eq:YukawaA2HDM}
\mathcal{L}_Y  &= - \frac{\sqrt{2}}{v}\, H^+\, \Big\{\bar{u} \left[  \varsigma_d \,V M_d P_R -  \varsigma_u \,M_u^{\dagger} V P_L\right] d \\
 & + \varsigma_\ell \,\bar{\nu} M_\ell P_R \ell \Big\} - \frac{1}{v}\; \sum_{\varphi^0_i,f}\, y^{\varphi^0_i}_f\, \varphi^0_i \, \left[\bar{f} M_f P_R f \right] + \mathrm{h.c.} \,,\nn
\end{align}
where $P_{R,L}\equiv \frac{1\pm \gamma_5}{2}$, $V$ is the CKM quark-mixing matrix
and the neutral Yukawa couplings are given by
%. The couplings of the neutral scalar fields to fermion pairs are given by
\begin{align} \label{eq:yukascal}
y_{d,\ell}^{\varphi^0_i} &= \cR_{i1} + (\cR_{i2} + i\,\cR_{i3})\,\varsigma_{d,\ell} \,,\\
y_u^{\varphi^0_i} &= \cR_{i1} + (\cR_{i2} - i\,\cR_{i3})\,\varsigma_{u}^* \,.
\end{align}
The usual $Z_2$ symmetric models can be recovered with specific assignments of the alignment parameters.
% All $\mathcal{Z}_2$ symmetric models can be recovered by A2HDM with specific assignment of the alignment parameters. We will discuss this in the section \ref{sec:z2symmetrycase}.

\subsection{Flavour misalignment}

The alignment conditions~(\ref{eq:alignment}) presumably hold at some high-energy scale $\Lambda_A$ and are spoiled by radiative corrections which induce a misalignment of the Yukawa matrices. However, the flavour symmetries of the A2HDM tightly constrain the possible FCNC structures, keeping their effects well below the present experimental bounds.
The only FCNC local structures induced at one loop take the form~\cite{Pich:2009sp,Jung:2010ik},
\begin{align} \label{eq:misalignment}
\mathcal{L}_{\mathrm{FCNC}} &= \frac{\cC}{4\pi^2 v^3} \left(1+\varsigma_u^*\,\varsigma_d\right)\\
& \hspace{-1.0cm} \times  \sum_i\, \varphi^0_i \Big\{(\cR_{i2} + i\,\cR_{i3})\, (\varsigma_d-\varsigma_u)\, \left[\bar{d}_L\, V^{\dagger} M_u M_u^\dagger V M_d\, d_R\right] \nn\\
& \hspace{-1.0cm} - (\cR_{i2} - i\,\cR_{i3})\, (\varsigma_d^*-\varsigma_u^*)\, \left[\bar{u}_L\, V M_d M_d^\dagger V^\dagger M_u\, u_R\right] \Big\} + \mathrm{h.c.}\, .\nn
\end{align}
The renormalization of the misalignment parameter $\cC$ is determined to be~\cite{Jung:2010ik}
\begin{equation} \label{eq:Crenorm}
\cC = \cC_R(\mu) + \frac{1}{2}\,\left\{\frac{2\mu^{D-4}}{D-4} +\gamma_E-\ln{(4\pi)}\right\}\, ,
\end{equation}
and absorbs the UV divergences from one-loop Higgs-penguin diagrams in $\Bqll$ decays~\cite{Li:2014fea}.

\section{Effective Hamiltonian}
\label{sec:Heff}

The low-energy effective Hamiltonian describing $\Bqll$ decays is given by~\cite{Buras:2013uqa,DeBruyn:2012wk,Altmannshofer:2011gn}
\begin{align} \label{eq:Heff}
 {\cal H}_{\rm eff}  &\, =\, -\frac{G_F \alpha}{\sqrt{2}\pi s^2_W}\,\left[V_{tb}V_{tq}^*\,\sum_{i}^{10,S,P} C_i\, {\cal O}_i + \mathrm{h.c.}\right]\,,\nn\\[0.2cm]
{\cal O}_{10} &\, =\, (\bar q\gamma_\mu P_L b)\, (\bar \ell \gamma^\mu \gamma_5 \ell)\,, \nn\\
{\cal O}_{S} &\, =\, \frac{m_\ell m_b}{M^2_W}\; (\bar q P_R b)\, (\bar \ell\ell)\,, \nn\\
{\cal O}_{P} &\, =\, \frac{m_\ell m_b}{M^2_W}\; (\bar q P_R b)\, (\bar\ell \gamma_5 \ell)\,,
\end{align}
where $\ell=e,\mu,\tau$; $q=d,s$, and $m_b=m_b(\mu)$ denotes the $b$-quark $\mathrm{\overline{MS}}$ running mass. Other possible operators are neglected because their contributions are either zero or
%they either have zero contribution or are
proportional to the light-quark mass $m_{q}$.

The anomalous dimension of
%the operator
${\cal O}_{10}$ is zero due to the conservation of the $(V-A)$ quark current in the massless quark limit. The operators ${\cal O}_{S}$ and ${\cal O}_{P}$ also have zero anomalous dimensions because the $\mu$ dependences
% anomalous dimensions
of $m_b(\mu)$ and the scalar current $(\bar q P_R b)(\mu)$ cancel each other. Therefore the Wilson coefficients $C_i$ do not receive additional renormalization from QCD corrections.

\section{Calculation of the Wilson coefficients $C_{10,S,P}$}

The Wilson coefficients $C_{10,S,P}$ are obtained by requiring the equality of one-particle irreducible amputated Green functions in the full and in the effective theories. The relevant Feynman diagrams for a given process can be created by the package \texttt{FeynArts}~\cite{Hahn:2000kx}, with the model files provided by \texttt{FeynRules}~\cite{Christensen:2008py}. The generated decay amplitudes are evaluated either with the help of \texttt{FeynCalc}~\cite{Mertig:1990an}, or using standard techniques such as the Feynman parametrization to combine propagators. We found full agreement between the results obtained with these two methods. Throughout the whole calculation, we set the light-quark masses $m_{d,s}$ to zero; while for $m_b$, we keep it up to linear order.

In general, the Wilson coefficients $C_i$ are functions of the internal up-type quark masses, together with the corresponding CKM factors~\cite{Buchalla:1995vs}:
\begin{equation}
C_i\; =\; \sum_{j=u,c,t}V_{jq}^* V_{jb}^{\phantom{*}}\; F_i(x_j)\, ,
\end{equation}
where $x_j=m_j^2/M_W^2$, and $F_i(x_j)$ denote the loop functions. In deriving the effective Hamiltonian (\ref{eq:Heff}), the limit $m_{u,c}\to 0$ and the unitarity of the CKM matrix,
\begin{equation} \label{eq:ckmunitary}
V_{uq}^* V_{ub}^{\phantom{*}} + V_{cq}^* V_{cb}^{\phantom{*}} + V_{tq}^* V_{tb}^{\phantom{*}}\; =\; 0\, ,
\end{equation}
have to be exploited. This implies that we need only to calculate explicitly the contributions from internal top quarks, while those from up and charm quarks are taken into account by means of simply omitting the mass-independent terms in the basic functions $F_i(x_t)$.

The relevant Feynman diagrams are split into various box, penguin and self-energy diagrams, which are mediated by the top quark, gauge bosons, and Higgs scalars. In order to check the gauge independence of the final results, we perform the calculation both in the Feynman~($\xi=1$) and in the unitary~($\xi=\infty$) gauges.

\subsection{Wilson coefficients in the SM}

In the SM, the dominant contribution to the decays $\Bqll$ comes from the Wilson coefficient $C_{10}$, which arises from $W$-box and $Z$-penguin diagrams:
%. The Wilson coefficient $C_{10}$ can be written as
\begin{equation}
 C^{\rm{SM}}_{10} \;=\; - \eta_Y^{\rm EW}\,\eta_Y^{\rm QCD}\,Y_0(x_t)\,,
 \label{eq:C10SM}
\end{equation}
where
\begin{equation} \label{eq:Y0}
Y_0(x_t) \, =\, \frac{x_t}{8}\left[ \frac{x_t-4}{x_t-1} + \frac{3x_t}{(x_t-1)^2}\ln x_t \right]
\end{equation}
is the one-loop Inami-Lim function~\cite{Inami:1980fz}. The factors $\eta_Y^{\rm EW}$ and $\eta_Y^{\rm QCD}$ account for the NLO electroweak~\cite{Bobeth:2013tba} and the NNLO QCD corrections~\cite{Hermann:2013kca}, respectively.

The coefficients $C_S$ and $C_P$ receive SM contributions from box, $Z$ penguin, Goldstone-boson (GB) penguin and Higgs (h) penguin diagrams:
\begin{align} \label{SMgauge}
C^{\rm SM}_{S} &\, =\, C^{\rm box,\, \rm SM}_{S}
+C^{\rm h\, penguin,\, \rm SM}_{S} \, ,\\
 C^{\rm SM}_{P} &\, =\,
 C^{\rm box,\, \rm SM}_{P} +C^{\rm Z\, penguin,\, \rm SM}_{P} +C^{\rm GB\, penguin,\, \rm SM}_{P} \, .
\end{align}
The Goldstone contribution is of course absent in the unitary gauge. Explicit expressions can be found in~\cite{Li:2014fea}.

%The SM contributions to $C_S$ and $C_P$ are given as
%\begin{align} \label{SMgauge}
%C^{\rm SM}_{S} &= C^{\rm box,\, \rm SM}_{S, \,\rm Feynman}
%+C^{\rm h\, penguin,\, \rm SM}_{S, \rm Feynman} \nn \\
%& = C^{\rm box,\, \rm SM}_{S, \,\rm Unitary}+C^{\rm h\, penguin,\, \rm SM}_{S, \,\rm Unitary}\,,\\[0.2cm]
% C^{\rm SM}_{P} &=
% C^{\rm box,\, \rm SM}_{P, \,\rm Feynman} +C^{\rm Z\, penguin,\, \rm SM}_{P, \rm Feynman} +C^{\rm GB\, penguin,\, \rm SM}_{P, \,\rm Feynman} \nn\\
%& = C^{\rm box,\, \rm SM}_{P, \rm Unitary} +C^{\rm Z\, penguin,\, \rm SM}_{P, \rm
%\, Unitary}\,,
%\end{align}
%where $ C^{\rm box,\, \rm SM}$, $C^{\rm Z\, penguin,\, \rm SM}$, $C^{\rm GB\, penguin,\, \rm SM}$, $C^{\rm h\, penguin,\, \rm SM}$ denote the SM contributions from box, $Z$-penguin, Goldstone-boson-penguin, Higgs-penguin diagrams, respectively.

\subsection{Wilson coefficients in the A2HDM}
\label{sec:WCA2HDM}

In the A2HDM there are additional contributions from box and $Z$ penguin diagrams, involving $H^\pm$ exchanges, and from Higgs penguin diagrams. The only new contribution to $C_{10}$ comes from $Z$ penguin diagrams and is gauge independent by itself:
\begin{equation}
C^{\rm A2HDM}_{10} =C^{\rm Z\, penguin,\, \rm A2HDM}_{10}\, .
\end{equation}

%The A2HDM contributions to $C_{10,S,P}$ can be divided into two parts, one from the box and $Z$-penguin diagrams involving the charged Higgs bosons, and the other one from the Higgs-penguin diagrams.
%
%\subsubsection{Charged Higgs contribution}
%
%In the A2HDM, the only new contribution to $C_{10}$ comes from the $Z$-penguin diagrams, which is gauge independent by itself:
%\begin{equation}
%C^{\rm A2HDM}_{10} =C^{\rm Z\, penguin,\, \rm A2HDM}_{10}
%\end{equation}

The $Z$ penguin diagrams also generate contributions to $C_P$. The sum of $Z$ penguin diagrams and Goldstone-boson penguin diagrams is gauge independent:
\begin{equation}
C^{\rm Z\, penguin,\, \rm A2HDM}_{P, \,\rm Unitary} =  C^{\rm Z\, penguin,\, \rm A2HDM}_{P, \,\rm Feynman}+C^{\rm GB\, penguin,\, \rm A2HDM}_{P, \,\rm Feynman}\, .
\end{equation}
The contributions from box diagrams with $H^\pm$ bosons, $C^{\rm box,\, \rm A2HDM}_{S,P}$, % and $C^{\rm box,\, \rm A2HDM}_P$,
are gauge dependent.
%, which will be cancelled by the ones from the neutral scalar exchanges.
\goodbreak
%\subsubsection{Neutral scalar exchange}

Neutral scalar exchanges induce both tree and loop diagrams. The loop contributions consist of the Higgs-penguin and self-energy diagrams governed by the Yukawa couplings~(\ref{eq:YukawaA2HDM}), whereas the tree ones are given by the misalignment couplings~(\ref{eq:misalignment}).
% that are suppressed by $m_{q}m_{q'}^2/v^3$ and quark-mixing factors.
The sum of these contributions can be written as:
\begin{align} \label{eq:ScalarPenguinWCs}
C_S^{\varphi_i^0,\, \rm A2HDM} &\, =\, \sum_{\varphi_i^0}\; \mathrm{Re} (y_\ell^{\varphi^0_i})
\; \hat{C}^{\varphi_i^0}\, ,\\
C_P^{\varphi_i^0,\, \rm A2HDM} &\, =\, i\; \sum_{\varphi_i^0}\; \mathrm{Im} (y_\ell^{\varphi^0_i})
\; \hat{C}^{\varphi_i^0}\,,
\end{align}
%where $\hat{C}^{\varphi_i^0}$ is given by
with
%\begin{align} \label{eq:chat}
%\hat{C}^{\varphi_i^0} &= x_t\,\Biggl\{
%\frac{(\varsigma_u-\varsigma_d)\, (1+\varsigma^*_u\varsigma_d)}{2 x_{\varphi^0_i}}\; (\cR_{i2} + i \cR_{i3})
%\; \cC_R(M_W)\nn\\
%& +\frac{v^2}{M^2_{\varphi^0_i}}  \lambda^{\varphi_i^0}_{H^+H^-}\;
%g_0^{\phantom{()}}(x_t,x_{H^+},\varsigma_u,\varsigma_d) + \sum_{j=1}^3 %\hspace{-0.1cm}
%\cR_{ij} \xi_j\,  \bigg[ \frac{1}{2 x_{\varphi^0_i}}\;\;\; \nn \\
%& \times g_j^{(a)}(x_t,x_{H^+},\varsigma_u,\varsigma_d)  + g_j^{(b)}(x_t,x_{H^+},\varsigma_u,\varsigma_d)\bigg] %\hspace{-0.07cm}
%\Biggr\}\,,
%\end{align}
%
\begin{align} \label{eq:chat}
\hat{C}^{\varphi_i^0} &= x_t\,\Biggl\{
\frac{(\varsigma_u-\varsigma_d)\, (1+\varsigma^*_u\varsigma_d)}{2 x_{\varphi^0_i}}\; (\cR_{i2} + i \cR_{i3})
\; \cC_R(M_W)\nn\\
& +\frac{v^2}{M^2_{\varphi^0_i}}  \lambda^{\varphi_i^0}_{H^+H^-}\;
g_0^{\phantom{()}} %(x_t,x_{H^+},\varsigma_u,\varsigma_d)
+ \sum_{j=1}^3 %\hspace{-0.1cm}
\cR_{ij} \xi_j\,  \bigg[ \frac{g_j^{(a)}}{2 x_{\varphi^0_i}}
 + g_j^{(b)}\bigg] %\hspace{-0.07cm}
\Biggr\}\,,
\end{align}
where $\lambda^{\varphi_i^0}_{H^+H^-} = \lambda_3\,\cR_{i1} + \lambda_7^R\,\cR_{i2}-\lambda_7^I\,\cR_{i3}$,
%is the rescaled cubic coupling $\varphi_i^0H^+G^-$,
$\xi_1=\xi_2 = 1$ and $\xi_3=i$. When $\varsigma_{u,d}\to 0$, $x_{H,A}\to\infty$, $x_h\to x_{h_{\rm SM}}$, $\cR_{i2,i3}\to 0$ and $\cR_{11}\to1$, this reproduces the SM result.

The coefficients $g_0^{\phantom{()}}$, $g_{j}^{(a)}$ and $g_{j}^{(b)}$ are functions of
$x_t$, $x_{H^+}$, $\varsigma_u$ and $\varsigma_d$. $g_0^{\phantom{()}}$ and $g_{j}^{(a)}$
%
%The functions $g_0^{\phantom{()}}(x_t,x_{H^+},\varsigma_u,\varsigma_d)$ and $g_{j}^{(a)}(x_t,x_{H^+},\varsigma_u,\varsigma_d)$
are gauge independent because they do not involve any gauge bosons,
while $g_{j}^{(b)}$ are all related to Goldstone-boson vertices and, therefore, are identically zero in the unitary gauge. These $g_{j}^{(b)}$ contributions cancel the gauge dependence from the box diagrams:
%In the unitary gauge, $g_{j}^{(b)}(x_t,x_{H^+},\varsigma_u,\varsigma_d)$ are zero, since they are all related to the Goldstone-boson vertices.
% The gauge dependence from the box diagrams are cancelled by  $g_{j}^{(b)}(x_t,x_{H^+},\varsigma_u,\varsigma_d)$,
\begin{eqnarray} \label{eq:CSSM-gauge}
C^{\rm box,\, \rm SM}_{S, \,\rm Unitary} \; - \;  C^{\rm box,\, \rm SM}_{S, \,\rm Feynman}
& = & x_t\, g_1^{(b)}  , \\[0.2cm]
      \label{eq:CSA2HDM-gauge}
C^{\rm box,\, \rm A2HDM}_{S, \,\rm Unitary} \; - \;  C^{\rm box,\, \rm A2HDM}_{S, \,\rm Feynman} &  = & \nn\\
&&\hspace{-3cm}  x_t\left[\mathrm{Re}(\varsigma_\ell) \; g_2^{(b)} - i\,\mathrm{Im}(\varsigma_\ell) \; g_3^{(b)}\right]  ,\\[0.2cm]
       \label{eq:CPA2HDM-gauge}
C^{\rm box,\, \rm A2HDM}_{P, \,\rm Unitary} \; - \;  C^{\rm box,\, \rm A2HDM}_{P, \,\rm Feynman} & = & \nn\\
&&\hspace{-3cm} x_t\left[ i\,\mathrm{Im}(\varsigma_\ell)\; g_2^{(b)} - \mathrm{Re}(\varsigma_\ell)\; g_3^{(b)} \right] .
\end{eqnarray}

The loop contributions with neutral scalar exchanges generate UV divergences which are cancelled by the renormalization of the misalignment coupling in (\ref{eq:Crenorm}). The $\mu$ dependence of the results is reabsorbed into the combination $\cC_R(M_W) = \cC_R(\mu) -\ln{(M_W/\mu)}$.
% Detailed expressions for all contributions can be found in~\cite{Li:2014fea}.

%The loop-induced contributions due to neutral scalar exchanges are not finite, and the UV divergence is cancelled by the renormalization of the misalignment coupling $\cC$ in (\ref{eq:misalignment}), and the $\mu$ dependence of the results are reabsorbed into the combination $\cC_R(M_W) = \cC_R(\mu) -\ln{(M_W/\mu)}$.

\section{Phenomenological analysis}

Currently, only $B_s\to \mu^+\mu^-$ is observed with a signal significance of $\sim 4.0\,\sigma$~\cite{CMSandLHCbCollaborations:2013pla}. Thus we shall investigate the allowed parameter space of the A2HDM under the constraint from $\overline{\mathcal{B}}(\Bsmm)$.
%
%\subsection{Branching ratio of $\Bsmm$}
%\label{sec:BR}
%
With updated input parameters, the SM prediction
% for $\Bsmm$
reads
\begin{equation} \label{eq:BsmmSM}
\overline{\mathcal{B}}(\Bsmm)_{\rm SM}  = (3.67 \pm 0.25) \times 10^{-9}\,.
\end{equation}
In order to explore constraints on the model parameters, it is convenient to introduce the ratio~\cite{Buras:2013uqa,DeBruyn:2012wk}
\begin{equation} \label{eq:R_bar_CPcons}
 \overline{R}_{s\ell}  \,\equiv\,  \frac{\overline{\mathcal{B}}(B_s^0\to \ell^+\ell^-)}{\overline{\mathcal{B}}(B_s^0\to \ell^+\ell^-)_{\rm SM}}
\, =\, \bigg[\,|P|^2+\Big(1-\frac{\Delta\Gamma_s}{\Gamma^s_L}\,\Big)\, |S|^2\bigg]\,,
\end{equation}
where $\Gamma^{s}_{L(H)}$ denote the lighter~(heavier) eigenstate decay width of the $B_s$ meson, and $\Delta\Gamma_s=\Gamma^{s}_L-\Gamma^{s}_H$.
% the width difference.
The quantities $S$ and $P$ are defined as
%, respectively, as
\begin{eqnarray} \label{eq:P}
\hspace{-0.3cm}  P &\!\!\!\equiv&\!\!\!  \frac{C_{10}}{C_{10}^{\rm SM}} + \frac{M^2_{B_s}}{2M^2_W} \left(\frac{m_b}{m_b+m_s}\right)\,\frac{C_P - C_{P}^{\rm SM}}{ C_{10}^{\rm SM}}\,, \\[0.2cm]
\label{eq:S}
\hspace{-0.3cm} S &\!\!\!\equiv &\!\!\! \sqrt{1-\frac{4m^2_\ell}{M^2_{B_s}}}\; \frac{M^2_{B_s}}{2M^2_W} \left(\frac{m_b}{m_b+m_s}\right)\,\frac{C_S - C_{S}^{\rm SM}}{ C_{10}^{\rm SM}} \,,
\end{eqnarray}
where the Wilson coefficients are given by:
% where the sum of the Wilson coefficients $C_{10}$, $C_{S}$ and $C_P$ are given by
\begin{eqnarray}\hspace{-0.3cm}
C_{10} &\!\!\! = &\!\!\! C^{\rm SM}_{10} +  C^{\rm Z\, penguin,\, \rm A2HDM}_{10} \,,\nn
\\[0.2cm]
C_{S} &\!\!\! =&\!\!\!  C^{\rm box,\, \rm SM}_{S}+  C^{\rm box,\, \rm A2HDM}_{S}
+ C^{\varphi_i^0,\, \rm A2HDM}_{S} \,,\nn \\[0.2cm]
C_{P} &\!\!\! =&\!\!\!
C^{\rm SM}_{P} + C^{\rm box,\, \rm A2HDM}_{P} + C^{\varphi_i^0,\, \rm A2HDM}_{P}
\nn \\ &\!\!\! + &\!\!\!
C^{\rm Z\, penguin,\, \rm A2HDM}_{P} + C^{\rm GB\, penguin,\, \rm A2HDM}_{P}\, .
%C^{\rm box,\, \rm SM}_{P}+ C^{\rm Z\, penguin,\, \rm SM}_{P}\,+ C^{\rm GB\, penguin,\, \rm SM}_{P}\nn \\
%&&  +\; C^{\rm Z\, penguin,\, \rm A2HDM}_{P}+ C^{\rm GB\, penguin,\, \rm A2HDM}_{P}
% \nn \\
%&& +\;  C^{\rm box,\, \rm A2HDM}_{P}+\, C^{\varphi_i^0,\, \rm A2HDM}_{P}\,.
\end{eqnarray}
Combining the SM prediction~(\ref{eq:BsmmSM}) with the latest experimental result (\ref{eq:Bsmmexp}), we get
\begin{equation} \label{eq:R_bar_results}
 \overline{R}_{s\mu} = 0.79 \pm 0.20\,.
\end{equation}

\subsection{Model parameters}

%In this work, we only consider the CP-conserving case in the Higgs sector,
We consider the CP-conserving limit and assume that the lightest CP-even scalar $h$ corresponds to the observed neutral boson with $M_h\simeq126~\mathrm{GeV}$. We have then 10 free parameters: 3 alignment couplings $\varsigma_f$, 3 scalar masses~($M_H$, $M_A$, $M_{H^\pm}$), 2 scalar-potential couplings~($\lambda_3$, $\lambda_7$), the mixing angle $\tilde\alpha$ and the misalignment parameter $C_R(M_W)$.
Four of them ($\tilde\alpha$, $\lambda_{3,7}$, $ C_R(M_W)$) have minor impacts on
$\overline{R}_{s\mu}$,
%the branching ratio
compared to the others. In order to simplify the analysis, we assign them the following values, using the bounds from earlier studies:
\begin{equation}
\lambda_3=\lambda_7 = 1, \quad \cos\tilde\alpha = 0.95, \quad C_R(M_W) =0 \,.
\end{equation}

The $C_{S,P}$ contributions to (\ref{eq:P}) and (\ref{eq:S}) are suppressed by a factor $M^2_{B_s}/M^2_W$.
% , compared to $C_{10}$.
Therefore, unless there are large enhancements from the $\varsigma_f$
% alignment
parameters, the branching ratio shall be dominated by $C_{10}$, where the A2HDM
% new physics~(NP)
contribution depends only on $|\varsigma_u|^2$ and $M_{H^\pm}$.
%the charged-scalar mass.
%
%From (\ref{eq:P}) and (\ref{eq:S}), we can see that the Wilson coefficients $C_S$ and $C_P$ are suppressed by the factor $M^2_{B_q}/M^2_W$ compared to $C_{10}$. Therefore, unless there are large enhancement from the alignment parameters, the branching ratio shall be dominated by $C_{10}$, where the new physics~(NP) contribution depends only on $|\varsigma_u|^2$ and the charged-scalar mass.
%
We shall then discuss two possible scenarios: 1) $|\varsigma_{d,\ell}|\lesssim |\varsigma_u|\leq 2$, where $C_{10}$ dominates, and 2) $|\varsigma_{d,\ell}| \gg |\varsigma_u|$, where $C_S$ and $C_P$ could play a significant role.

\subsection{Small $\varsigma_{d,\ell}$}

%When the alignment parameters $\varsigma_{d,\ell}$ are close to the size of $\varsigma_u$, the NP contributions from $C_S$ and $C_P$ are negligible. The NP contribution to $C_{10}$ is $C^{\rm A2HDM}_{10}$, which involves only two free parameters, $\varsigma_u$ and $M_{H^\pm}$. Especially, $C^{\rm A2HDM}_{10}$ goes to zero when $\varsigma_u \to 0$ and/or $M_{H^\pm}\to\infty$.

\begin{figure}[t]
\centering
\includegraphics[width=\linewidth]{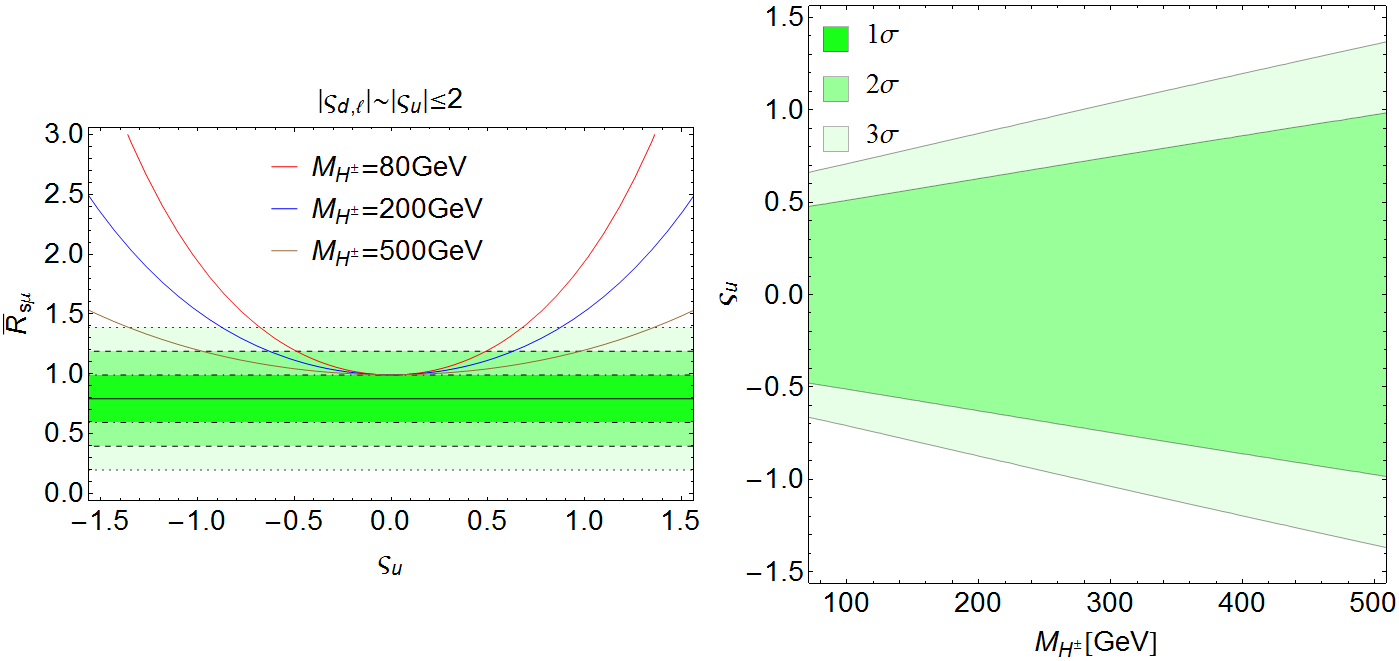}
\caption{\small Dependence of $\bar R_{s\mu}$ on $\varsigma_u$~(left)
%with three typical charged-Higgs masses~($80$, $200$ and $500~\mathrm{GeV}$), and the
and resulting upper bounds on $\varsigma_u$~(right), as function of $M_{H^\pm}$, for $|\varsigma_{d,\ell}|\lesssim |\varsigma_u|\leq 2$.}
\label{plot:C10}
\end{figure}

The only relevant NP contribution is $C^{\rm A2HDM}_{10}$ which involves two parameters, $\varsigma_u$ and $M_{H^\pm}$.
The constraints imposed by
% on the parameters $\varsigma_u$ and $M_{H^\pm}$ by
$\bar R_{s\mu}$ are shown in Fig.~\ref{plot:C10}. In the left panel, we choose $M_{H^\pm} = 80$, $200$ and $500~{\rm GeV}$~(upper, middle and lower curves, respectively). The shaded horizontal bands denote the allowed experimental region at $1\sigma$~(dark green), $2\sigma$~(green), and $3\sigma$~(light green), respectively. The right panel shows the resulting upper bounds on $\varsigma_u$, as function of $M_{H^\pm}$. A $95\%$ CL upper bound $|\varsigma_u|\leq 0.49~(0.97)$ is obtained for $M_{H^\pm}=80~(500)~\mathrm{GeV}$.
Since $C_{10}^{\rm A2HDM} \sim |\varsigma_u|^2$, this constraint is independent of any assumption about CP. For larger masses the constraint becomes weaker since the $H^\pm$ contribution starts to decouple.

%charged-Higgs masses, the constraint becomes weaker as the NP effect starts to decouple, reflected by $\underset{x_{H^+}\to \infty}{\lim}C_{10}^{\rm A2HDM}=0$.

%\vspace{0.5cm}
%\begin{minipage}{0.95\linewidth}
%\centering
%\includegraphics[width=\textwidth]{plotsmallzdzl.png}
%\captionof{figure}{\small The dependence of $\bar R_{s\mu}$ on the alignment parameter $\varsigma_u$~(left) with three typical charged-Higgs masses~($80$, $200$ and $500~\mathrm{GeV}$), and the resulting upper bounds on $\varsigma_u$, as function of $M_{H^\pm}$~(right).}
%\label{plot:C10}
%\end{minipage}
%\vspace{0.5cm}

\subsection{Large $\varsigma_{d,\ell}$}

In this case, $C_S$ and $C_P$ can induce a significant impact on the branching ratio. We vary $\varsigma_d$ and $\varsigma_\ell$ within the range $[-50, 50]$, and choose three representative values for $\varsigma_u=0,\pm1$. We also take three different representative sets of scalar masses:
\begin{align}
& \rm{Mass1}: \quad M_{H^\pm}= M_A = 80~\mathrm{GeV}, \quad M_H = 130~\mathrm{GeV}\,, \no \\
& \rm{Mass2}:  \quad M_{H^\pm}= M_A = M_H = 200~\mathrm{GeV}\,, \no\\
& \rm{Mass3}:  \quad M_{H^\pm}= M_A = M_H = 500~\mathrm{GeV}\,.
\end{align}

\begin{figure}[t]
\centering
\includegraphics[width=\linewidth]{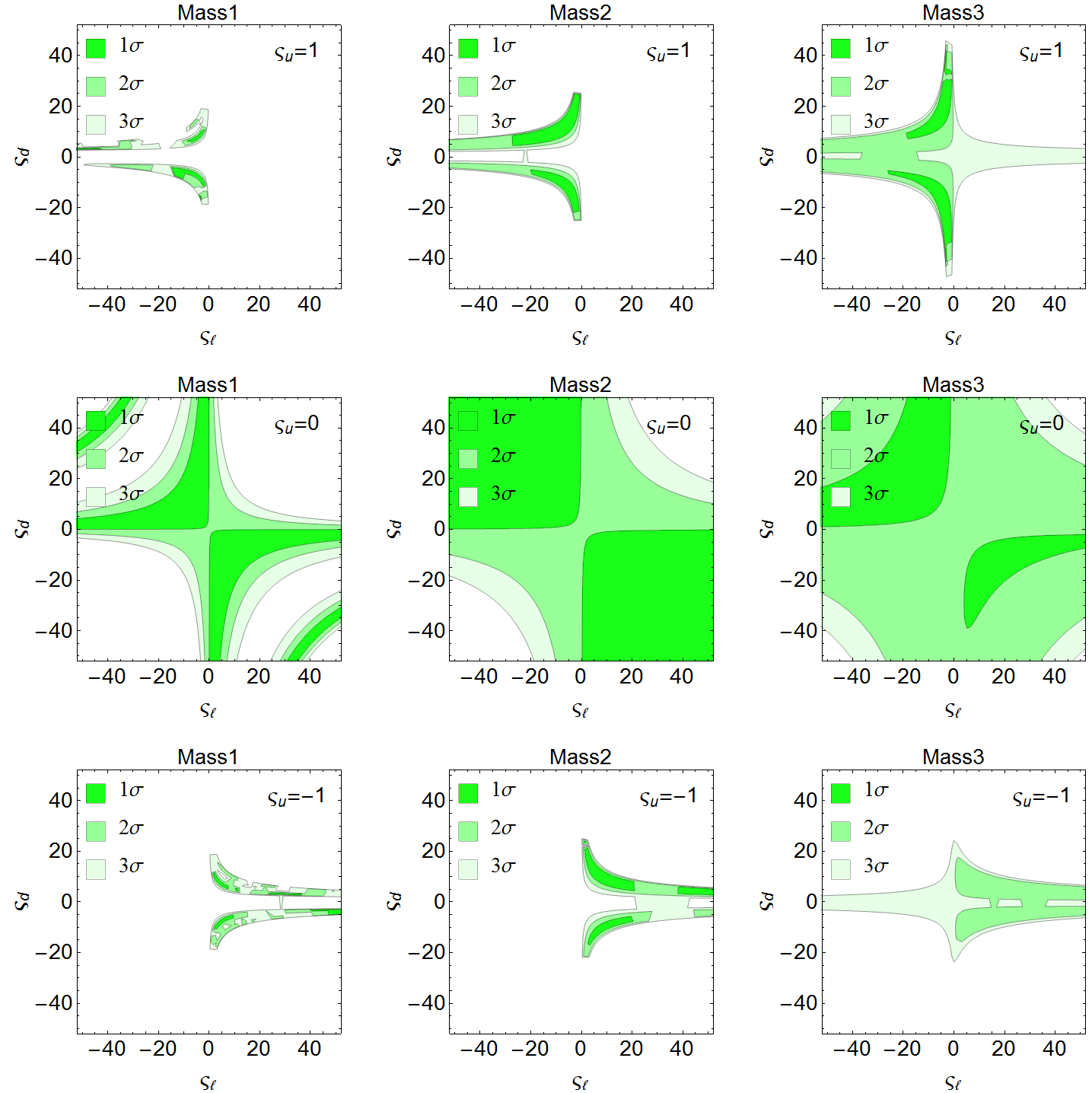}
\captionof{figure}{\small Allowed regions~(at $1\sigma$, $2\sigma$ and $3\sigma$) in the $\varsigma_d$--$\varsigma_\ell$ plane under the constraint from $\bar R_{s\mu}$, with three different assignments of the scalar masses and $\varsigma_u=0,\pm1$.}
\label{plot:zdzl}
\end{figure}

In Fig.~\ref{plot:zdzl}, we show the allowed regions in the $\varsigma_d$--$\varsigma_\ell$ plane under the constraint from $\bar R_{s\mu}$. The regions with large $\varsigma_d$ and $\varsigma_\ell$ are already excluded, especially when they have the same sign. The impact of $\varsigma_u$ is significant: a nonzero $\varsigma_u$ will exclude most of the regions allowed in the case with $\varsigma_u=0$, and changing the sign of $\varsigma_u$ will also flip that of $\varsigma_\ell$. The allowed regions expand with increasing scalar masses, as expected, since the NP contributions gradually decouple from the SM.

%\vspace{0.5cm}
%\begin{minipage}{0.95\linewidth}
%\centering
%\includegraphics[width=\textwidth]{plotlargezdzl.png}
%\captionof{figure}{\small Allowed regions~(at $1\sigma$, $2\sigma$ and $3\sigma$) in the $\varsigma_d$--$\varsigma_\ell$ plane under the constraint from $\bar R_{s\mu}$, with three different assignments of the scalar masses and $\varsigma_u=0,\pm1$.}
%\label{plot:zdzl}
%\end{minipage}

\subsection{2HDMs with discrete $Z_2$ symmetries}
\label{sec:z2symmetrycase}

The usual $Z_2$ symmetric models are recovered for the values of $\varsigma_f$ indicated in Table~\ref{tab:models}. In these models, the ratio $\bar R_{s\mu}$ only involves seven free parameters: $M_{H^\pm}$, $M_H$, $M_A$, $\lambda_3$, $\lambda_7$, $\cos\tilde\alpha$ and $\tan\beta$. In the particular case of the type-II 2HDM at large $\tan\beta$, our results agree with the ones calculated in Ref.~\cite{Logan:2000iv}. It is also interesting to note that the $\Bqll$ branching ratios depend only on the charged-Higgs mass and $\tan\beta$ in this case.

%%%%%%%%%%%%%%%%%%%%%%%%%%%%%%%%%%%%%%%%%%%%%%%%%%%%%%%%%%%%%%%%%%%
\begin{table}[t]
  \centering
  \tabcolsep 0.08in
  \begin{tabular}{|c|c|c|c|}
  \hline
  Model   & $\varsigma_d$  & $\varsigma_u$ & $\varsigma_l$  \\
  \hline
  Type~I  & $ \cot{\beta}$ & $\cot{\beta}$ & $ \cot{\beta}$ \\
  Type~II & $-\tan{\beta}$ & $\cot{\beta}$ & $-\tan{\beta}$ \\
  Type~X (lepton-specific) & $ \cot{\beta}$ & $\cot{\beta}$ & $-\tan{\beta}$ \\
  Type~Y  (flipped) & $-\tan{\beta}$ & $\cot{\beta}$ & $ \cot{\beta}$ \\
  Inert   &      0         &        0      &       0        \\
  \hline
  \end{tabular}
  \caption{\label{tab:models} \small 2HDMs based on discrete $Z_2$ symmetries.}
\end{table}
%%%%%%%%%%%%%%%%%%%%%%%%%%%%%%%%%%%%%%%%%%%%%%%%%%%%%%%%%%%%%%%%%%%

\begin{figure}[t]
\centering
\includegraphics[width=\linewidth]{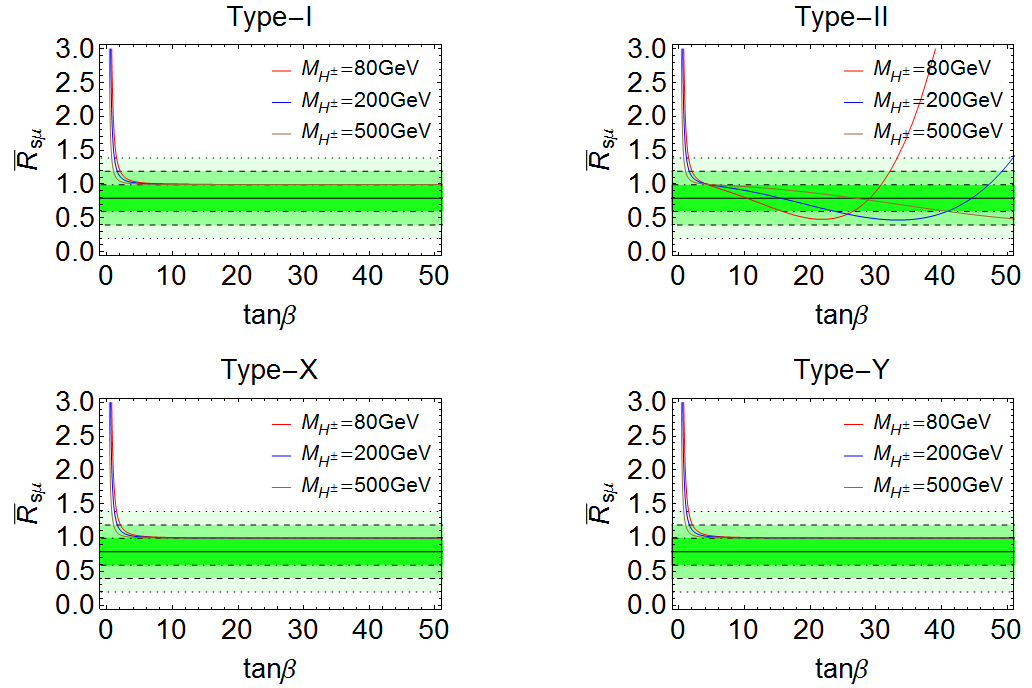}
\captionof{figure}{\small Dependence of $\bar R_{s\mu}$ on $\tan\beta$ for the 2HDMs of types I, II, X and Y. The upper, middle and lower curves correspond to $M_{H^\pm} = 80$, $200$ and $500~\mathrm{GeV}$, respectively. The horizontal bands denote the allowed experimental region at $1\sigma$~(dark green), $2\sigma$~(green), and $3\sigma$~(light green).}
\label{plot:z2types}
\end{figure}

Fig.~\ref{plot:z2types} shows the dependence of $\bar R_{s\mu}$ on $\tan\beta$, for three representative charged-Higgs masses: $M_{H^\pm}=80$, $200$ and $500~\mathrm{GeV}$. The other two neutral scalar masses have been fixed at $M_{H}=M_{A}=500~\mathrm{GeV}$. The four different panels correspond to the models of types I, II, X and Y, respectively. A lower bound $\tan\beta>1.6$ is obtained at $95\%$ CL under the constraint from the current experimental data on $\bar R_{s\mu}$. This implies $\varsigma_u = \cot{\beta} < 0.63$, which is stronger than the bounds obtained previously from other sources~\cite{Jung:2010ik,A2HDM:flavour}.

%\vspace{0.5cm}
%\begin{minipage}{0.95\linewidth}
%\centering
%\includegraphics[width=\textwidth]{plotz2symmetry.png}
%\captionof{figure}{\small Dependence of $\bar R_{s\mu}$ on $\tan\beta$ for the 2HDMs of types I, II, X and Y. The upper, middle and lower curves correspond to $M_{H^\pm} = 80$, $200$ and $500~\mathrm{GeV}$, respectively. The horizontal bands denote the allowed experimental region at $1\sigma$~(dark green), $2\sigma$~(green), and $3\sigma$~(light green), respectively.}
%\label{plot:z2types}
%\end{minipage}

\section{Conclusions}
\label{sec:conclusion}

We have studied the rare decays $\Bqll$ within the general framework of the A2HDM. A complete one-loop calculation of the Wilson coefficients $C_{10}$, $C_S$ and $C_P$ has been performed. They arise from various box, penguin and self-energy diagrams, as well as tree-level FCNC diagrams induced by the flavour misalignment interaction~(\ref{eq:misalignment}).
The gauge independence of the results has been checked through separate calculations in the Feynman and unitary gauges, and the gauge relations among different diagrams have been examined in detail.

We have also investigated the impact of the current $\overline{\mathcal{B}}(B_s^0\to \mu^+\mu^-)$ data on the model parameters, especially the resulting constraints on the three alignment couplings $\varsigma_f$. This information is complementary to the one obtained from collider physics and will be useful for future global data fits within the A2HDM.

%and constraints on the model parameters under the constraint from the current data on $\overline{\mathcal{B}}(B_s^0\to \mu^+\mu^-)$, especially on the three alignment parameters $\varsigma_f$.
%
%The resulting information about the model parameters will be useful for the model building and is complementary to the collider physics.

\section*{Acknowledgements}
Work supported by the NSFC~[contracts~11005032 and 11435003], the Spanish Government~[FPA2011-23778]
%  and ERDF funds from the EU Commission~[Grants No. FPA2011-23778, No. CSD2007-00042~(Consolider Project CPAN)]
and Generalitat Valenciana~[PROMETEOII/2013/007]. X. Li was also supported by the Scientific Research Foundation for the Returned Overseas Chinese Scholars, State Education Ministry. J. Lu is grateful for the hospitality of Center for Future High Energy Physics in Beijing.



%% References

%%\nocite{*}
%%\bibliographystyle{elsarticle-num}
%%\bibliography{martin}



\end{document}